\documentclass[doublecol]{epl2} 
\usepackage{amsmath,amssymb}

\usepackage{graphicx}
\usepackage{amsmath}
\usepackage{bm,color}

\def\6#1{{\underline{#1}}}
\def\m6#1{{\underline{#1}\,}}

\newdimen\Tdim
\def\ispan{{\setbox0=\hbox{i}%
\Tdim\ht0\advance\Tdim\dp0\rule[-\dp0]{0pt}{\Tdim}}}
\def\jspan{{\setbox0=\hbox{j}%
\Tdim\ht0\advance\Tdim\dp0\rule[-\dp0]{0pt}{\Tdim}}}
\def\Tspan#1{{\setbox0=\hbox{#1}%
\Tdim\ht0\advance\Tdim\dp0\advance\Tdim.55ex\rule[-\dp0]{0pt}{\Tdim}\box0}}

\def\be{\begin{eqnarray}}
\def\ben{\begin{eqnarray*}}
\def\ee{\end{eqnarray}}
\def\een{\end{eqnarray*}}

\def\=:{=\hspace{-.7em}\raisebox{1.1ex}{.}\hspace{.1em}\raisebox{-0.2ex}{.} }

\newcommand {\1}[1]{\frac{1}{#1}}

\newcommand {\beq}{\begin{eqnarray}}
\newcommand {\eeq}{\end{eqnarray}}
\newcommand {\non}{\nonumber\\}

\title{Vortex graphs as $N$-omers and ${\mathbb C}P^{N-1}$ Skyrmions \\
in $N$-component Bose-Einstein condensates}
\shorttitle{Title} 

\author{Minoru Eto\inst{1} \and Muneto Nitta\inst{2}}
\shortauthor{M. Eto \etal}

\institute{                    
  \inst{1} Department of Physics, Yamagata University, Yamagata 990-8560, Japan\\
  \inst{2} Department of Physics, and Research and Education Center for Natural 
Sciences, Keio University, Hiyoshi 4-1-1, Yokohama, Kanagawa 223-8521, Japan
}
\pacs{03.75.Lm}{First pacs description}
\pacs{03.75.Mn}{Second pacs description}

\abstract{
Stable vortex $N$-omers are constructed in 
coherently coupled $N$-component Bose-Einstein condensates.
We classify all possible $N$-omers in terms of the mathematical graph theory 
and numerically construct all graphs for $N=2,3,4$.
We also find that $N$-omers are well described as 
${\mathbb C}P^{N-1}$ skyrmions when 
inter-component and intra-component couplings are
$U(N)$ symmetric, and we evaluate their size dependence 
on the Rabi coupling.
}

\begin{document}

\maketitle

{\it Introduction} --- 
Exotic vortices in multi-component condensates 
have recently attracted considerable attention 
in the area of condensed matter physics; in this light, the topics of interest include
exotic superconductors, superfluid $^3$He \cite{Volovik:2003}, 
Bose-Einstein condensates (BECs) \cite{Pethickbook,Ueda}, exciton-polariton condensates, 
and nonlinear optics \cite{Desyatnikov}.
In particular, recent advances in realizing BECs in ultracold atomic 
gases have opened up new possibilities in quantum physics \cite{Pethickbook,Ueda};
of particular interest is the tuning of the s-wave scattering wavelength via a Feshbach resonance 
\cite{Thalhammer,Papp:2008,Tojo:2010}. 
Multi-component BECs can be realized when 
more than one hyperfine spin state is simultaneously populated 
\cite{Myatt:1997,Tojo:2010}
or when more than one species of atoms are mixed 
\cite{Modugno:2002,Papp:2008,McCarron:2011}.
The recent experimental achievement of a condensate of 
ytterbium offers 
condensations up to five components \cite{Fukuhara:2007}. 
Vortices in multi-component BECs have been realized experimentally \cite{Matthews,Schweikhard:2004}, 
and the structures of these vortices are considerably  
richer than those formed of single components 
 \cite{Ho:1998,Semenoff:2006vv,
Turner:2009,
Mueller:2002,Kasamatsu:2003,Kasamatsu:2004,
Kasamatsu:2005,Kasamatsu:2009,
Ji:2008,Eto:2011,
Aftalion:2011,Aftalion:2012,Kuopanportti:2012,Eto:2012rc,
Cipriani:2013nya,Cipriani:2013wia}; 
one fascinating feature in such cases is that  
vortices are fractionally quantized.
The study of BECs can provide an ideal opportunity to examine the dynamics of exotic vortices because these vortices can theoretically be explained via 
a quantitative description using the mean field theory 
by the Gross-Pitaevski (GP) equation
and further, the vortices can be experimentally controlled to a large degree.

Vortices or fluxes are fractionally quantized 
in multi-gap superconductors
\cite{Tanaka:2001,Babaev:2002,Gurevich:2003,Goryo:2007} 
as in the case of multi-component BECs. 
One added feature of multi-gap superconductors 
is the presence of 
Josephson coupling between the gaps. 
Due to the presence of this coupling 
there appears a sine-Gordon soliton  
for the phase difference between the two gaps 
\cite{Tanaka:2001,Bluhm:2006}. 
When a set of  two vortices 
in two different gaps is separated, 
a sine-Gordon domain wall connects the vortices, thereby resulting in 
a two-vortex molecule, {\it i.e.}, a dimer \cite{Goryo:2007}. 
Multi-components are also present 
in thin-layer Josephson junctions and 
high-$T_{\rm c}$ superconductors 
with multi-layers. 
Recently discovered iron-based superconductors 
may have three gaps, in which case 
there may be three-vortex molecule {\it i.e.}, 
a trimer \cite{Nitta:2010yf}. 
However,  
the vortex molecules have not been observed directly 
thus far in superconductors except indirectly 
\cite{fractional-exp},  
because the repulsion between vortices is exponentially 
suppressed by the Meissner effect 
so that it cannot be balanced with the domain wall tension. 

In the case of multiple hyperfine spin states of BECs, 
internal coherent (Rabi) couplings between multiple components 
can be introduced by Rabi oscillations, 
similar to Josephson couplings in 
multi-gap superconductors. 
As in superconductors, 
a sine-Gordon domain wall of the phase difference of two components appears \cite{Son:2001td}. 
In this light, the advantages of BECs are that 
Rabi coupling can be tuned experimentally and that
a vortex dimer can exist stably 
because repulsion between 
two fractional vortices \cite{Eto:2011} 
is sufficiently strong  
in the absence of the Meissner effect,
to be balanced with the tension of 
a sine-Gordon domain wall connecting them 
\cite{Kasamatsu:2004}.
Recently, we observed stable-vortex trimers 
in three-component BECs, and 
we showed that the shape and the size of the molecule 
can be controlled by varying the strength of  
the Rabi couplings \cite{Eto:2012rc}. 

In this Letter, we construct stable-vortex $N$-omers, 
that is, molecules made of $N$ fractional vortices in 
$N$-component BECs,
and we find that $N$-omers with $N \ge 3$ 
exhibit several novel properties that dimers do not possess, {\it i.e.}, the existence of chirality pairs 
and 
metastable states such as twist, holding, and 
capture, which are properties exhibited by 
chemical molecules.
Each condensate wave function has 
a nontrivial winding number around one of 
the $N$ fractional vortices.
When the Rabi couplings
are turned on, 
the fractional vortices are glued by domain walls
to form a vortex $N$-omer.
We make use of the mathematical graph theory to
classify the vortex $N$-omers 
whose number exponentially increases as $N$ increases.
We numerically construct all possible graphs 
for $N=3,4$ by imaginary time propagation
with the phase winding and a constant density 
fixed by the Dirichlet conditions at the boundary, 
thereby our results
imply that the vortex $N$-omers are formed in rotating BECs. 
We find that two geometrically different graphs 
topologically equivalent to each other 
can both be stable. 
Furthermore, we can control the shape and size of the molecules  
by varying the strength of the Rabi couplings. 
To demonstrate this, we construct a vortex constellation,
{\it i.~e.}, 
the orion.
We also show that vortex $N$-omers are well described 
 by ${\mathbb C}P^{N-1}$ skyrmions when 
all possible Rabi couplings are set to 
the same values,  
where ${\mathbb C}P^{N-1} \simeq SU(N)/[SU(N-1)\times U(1)]$ 
denotes the complex projective space.

$N$-component BECs --- 
We consider coherently coupled $N$-component BECs of atoms with mass $m$, 
described by
the condensate wave functions $\Psi_i$ ($i=1,2,\cdots,N$) with the GP energy functional
\beq
E &=& \sum_{i,j=1}^N\int d^2x 
 \bigg(-\frac{\hbar^2}{2m}\Psi_i^*\nabla^2\Psi_i \delta_{ij}
+ \frac{g_{ij}}{2}|\Psi_i|^2 |\Psi_j|^2 \non
&&- 
\mu_i|\Psi_i|^2\delta_{ij} - \omega_{ij}\Psi_i^*\Psi_j
\bigg),
\label{eq:gp}
\eeq
where atom-atom interactions are characterized by 
the coupling constants $g_{ij} = 4\pi \hbar^2 a_{ij}/m$ 
with s-wave scattering lengths $a_{ij}$, 
$\mu_i$ denotes the chemical potential, and 
$\omega_{ij} = \omega_{ji}$ ($\omega_{ii} = 0$)
denotes the internal coherent coupling due to 
Rabi oscillations 
between the $i$-th and $j$-th components.
We mainly consider the case where 
$g_{ii} \equiv g$, 
 and $\mu_i \equiv \mu$,
and $g_{ij} \equiv \tilde g$ ($i\neq j$) in this Letter. 
It is straightforward to consider the general case but
all the results below are essentially unchanged from our simplest choice.

The symmetry of the Hamiltonian depends on the coupling constants $g,\tilde g$, and $\omega_{ij}$.
The maximal symmetry $U(N)$ is achieved when $g = \tilde g$ with $\omega_{ij} = 0$.
When $g \neq \tilde g$ and $\omega_{ij}=0$, 
this symmetry reduces to $U(1)^N$ 
acting on a phase of each component.
When $g \neq \tilde g$ and $\omega_{ij} \neq 0$, 
it further reduces to the gauge symmetry $U(1)$ 
which rotates the phases of all the components 
uniformly.

When all the internal coherent couplings are equal {\it i.e.},
$\omega_{ij}=\omega$, 
the condensations of the ground state are 
$|\Psi_i| = v \equiv \sqrt{\frac{\mu + (N-1)\omega}{g+(N-1)\tilde g}}$,
with $i=1,2,\cdots,N$. 
These amplitudes are modified  
in the case $\omega_{ij} \neq \omega_{i'j'}$, and
one should solve the variational problem $\delta E/ \delta \Psi_i = 0$ to determine $v_i = |\Psi_i|$ numerically.

As long as the internal coherent couplings $|\omega_{ij}|$ are maintained sufficiently small with respect to the other couplings
(we choose $g \sim \tilde g = {\cal O}(10^3)$, $\mu = {\cal O}(10^2)$, and $\omega = {\cal O}(10^{-2})$ 
in dimensionless units $\hbar = m =1$),
the symmetries $U(N)$ for $g= \tilde g$ and 
$U(1)^N$ for $g \neq \tilde g$ are nearly intact. 
These symmetries are spontaneously broken 
in the ground state. 
Consequently, vortices that are 
quite different for $g = \tilde g$ and $g > \tilde g$ can appear. 
For $g=\tilde g$, an axisymmetric giant vortex appears. 
As we will explain below, this can be interpreted as 
a $\mathbb{C}P^{N-1}$ skyrmion.  
On the other hand, for $g > \tilde g$, 
there appear $N$ fractional vortices associated with
the broken $U(1)^N$ symmetry,
which are connected by domain walls, thereby
resulting in molecules with $N$ vortices.

The internal coherent couplings 
$- \omega_{ij}\Psi_i^*\Psi_j = 
- 2v_iv_j\omega_{ij}\cos(\theta_i - \theta_j)$,
with $\theta_i = \arg \Psi_i$, 
glue the fractional vortices by creating 
domain walls among them. 
For $\omega_{ij} > 0$,  
all the phases $\theta_i$ and $\theta_j$ coincide, {\it i.e.}, $\theta_i = \theta_j$ in the ground state.

The typical scale of our system is $v^{-1} \sim \sqrt{\mu/g} \sim 3$ (in dimensionless unit). 
In all of our numerical simulations,
we set our box size about 10 times bigger than the typical scale, so that we safely ignore boundary effects.

First, 
let us consider the case that 
all $\omega_{ij}$'s are zero. 
A vortex winding around only one component, 
$\Psi_i = v_i \rho(r) e^{i\theta}$,  
in polar coordinates $r$ and $\theta$,   
with the other components almost constant, 
is fractionally quantized as
\beq
 \oint d {\bf r} \cdot {\bf v}_{\rm s} = {{v_i^2}\over {\sum_i} v_i^2} {h \over m},
\eeq
with the superfluid velocity ${\bf v}_{\rm s}$,
and it is stable \cite{Eto:2012rc}. 
However, once we turn on the internal coherent couplings $\omega_{ij}$, no matter how tiny they are, the vortex is inevitably attached by semi-infinitely long domain walls extending to the boundary 
because of the Rabi energy.
To examine this phenomenon,
we approximate 
the condensate wave functions by $|\Psi_i| = v_i e^{i \theta_i}$
to obtain the reduced energy functional 
${\cal E}_{\rm phase} = \sum_{i=1}^N\frac{\hbar^2}{2m}(\vec \nabla \theta_i)^2 
- 2\sum_{i>j}^N\omega_{ij}v_iv_j\cos (\theta_i-\theta_j)$.
This reduced energy functional shows that $N-1$ domain walls are attached 
to the $i$-th vortex in the direction $\theta_i = \theta_j$. 
Therefore, this vortex cannot remain stable in an isolated state, and it must be connected to the boundary or to vortices winding around the other components forming the molecules.

As described below, we find that the mathematical graph theory 
is useful to classify the vortex molecules, 
where we identify vertices as vortices, 
and the edge connecting $i$-th and $j$-th vertices 
as the Rabi coupling $\omega_{ij}$ if it is nonzero.

\begin{figure}
\begin{center}
\includegraphics[width=\hsize]{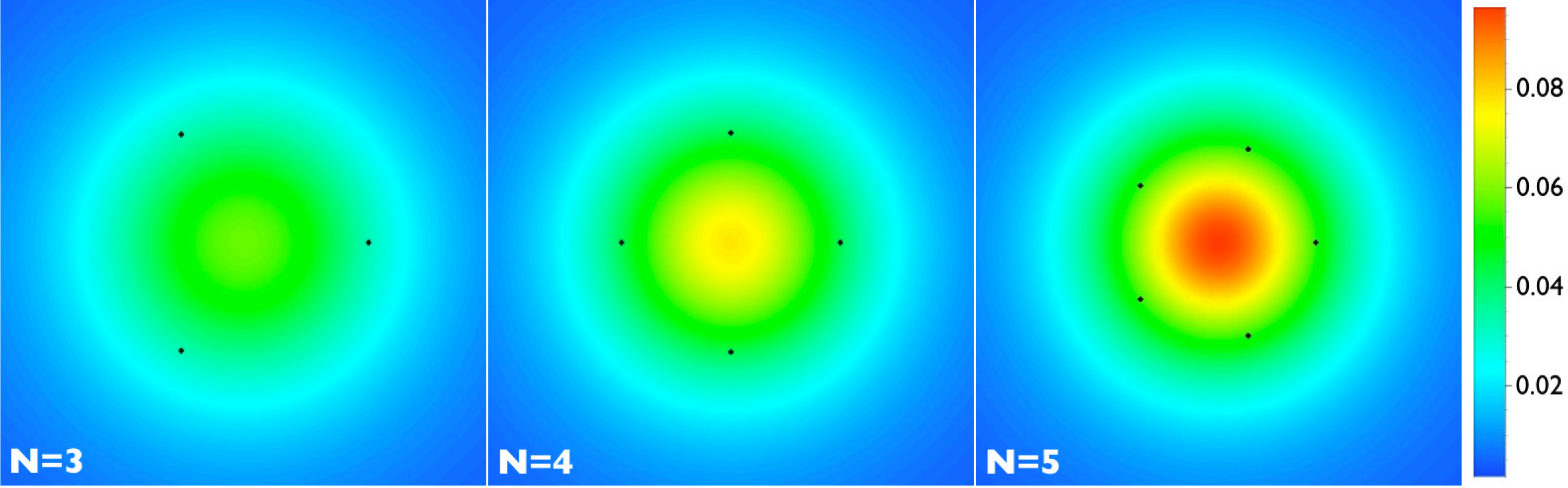}
\caption{Energy-density distribution of axisymmetric molecules
with $m=\hbar=1$, $g = \tilde g = 10^3$, $\mu = 10^2$, and $\omega = 0.05$.
The size of the simulation box is $(2L)^2 = 20^2$ with the lattice space set to be less than 0.1.
Zeros of $\Psi_i$ are indicated by the black dots. The numerical range for each of the figures 
is set as $x \in [-2.5,2.5]$ and $y \in [-2.5,2.5]$.}
\label{fig:axisymmetric}
\includegraphics[width=8.7cm]{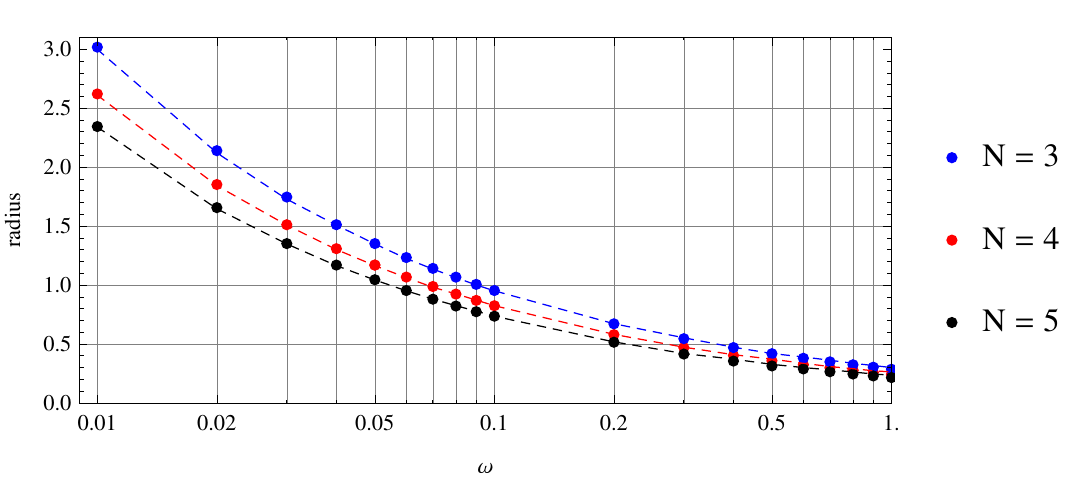}
\vspace*{-.5cm}
\caption{Sizes of $\mathbb{C}P^{N-1}$ skyrmions for $N=3,4,5$. 
The dots indicate skyrmion radii obtained from the numerical solutions. 
The radii obtained by the variational method are connected by broken lines.  
}
\label{fig:sizes}
\end{center}
\end{figure}
{\it $\mathbb{C}P^{N-1}$ skyrmions} ---
Here, we consider the symmetric case with $g = \tilde g$ and with all possible internal coherent couplings being the same  
($\omega_{ij} = \omega$ for all $i,j$), 
which correspond to the complete graphs.
We numerically solve the GP equation 
by using an imaginary time propagation 
for $N=3,4,5$, as shown in Fig.~\ref{fig:axisymmetric}.
We find that energy densities are always axisymmetric 
so that we may call them giant vortices. 
However, each component has a different configuration;
the vortex centers of each component are expressed 
as dots, whose positions are ${\mathbb Z}_N$ symmetric.  
We also varied $\omega$ and found that 
the size of the configuration decreases as 
$\omega$ increases, 
as shown in Fig.~\ref{fig:sizes} for $N=3,4,5$.
With our numerical survey over a wide range of parameter space and various initial configurations,
we conclude that
the existence of vortices is robust.

We further confirm the stability and existence of the solutions 
by using a variational method.
In the parameter region of $g \gg |\omega|$,
we note from Eq.~(\ref{eq:gp}) 
that  the total density is approximately constant, {\it i.e.},
$\sum_i |\Psi_i|^2 \simeq \mu/g$.
Thus, we are naturally led to define the reduced wave functions $\{\phi_i\}$ by
$\Psi_i = \sqrt{\frac{\mu}{Ng}}\, \phi_i,\quad \sum_{i=1}^N|\phi_i|^2 = 1$.
Consequently, the GP model expressed by Eq.~(\ref{eq:gp}) 
reduces to the $S^{2N-1}$ nonlinear sigma model.
Upon separating the total phase 
$\Theta \equiv \1{N}\sum_{i=1}^N \theta_i$,  
the rest of the degrees of freedom are 
described by the $\mathbb{C}P^{N-1}$ model,
\beq
{\cal E}_{{\mathbb C}P^{N-1}} \simeq \frac{\mu}{Ng}\left[\frac{\hbar^2}{2m} \sum_i \left|\nabla \phi_i\right|^2
- \omega_{ij} \left(\phi_i^*\phi_j + \phi_i\phi_j^*\right) 
 \right],
\label{eq:ene_reduce}
\eeq
with the identification $\phi_i \sim e^{i \alpha}\phi_i$. 
Motivated by exact solutions of ${\mathbb C}P^{N-1}$ skyrmions
and the ${\mathbb Z}_N$ symmetric zeros in Fig.~\ref{fig:axisymmetric}, 
we propose a variational ansatz 
\beq
&& \phi_i 
 = \frac{z - z_i }
           {\sqrt{\sum_{j=1}^N \left|z-z_j\right|^2}}, 
\quad z \equiv x+iy ,\\
&& z_j \equiv x_j + iy_j = \omega_N^{j-1}\lambda e^{-\frac{a}{2}r^2}, 
 \quad
 \omega_N \equiv e^{\frac{2\pi i}{N}}, \non
&& \Theta = {1 \over N} \sum_{i=1}^N \tan^{-1} 
{y-y_i \over x-x_i}
\eeq 
with variational parameters $a$ and $\lambda$, 
as a generalization of 
the case of $N=2$ (${\mathbb C}P^1$)
\cite{Kasamatsu:2004, Kasamatsu:2005}. 
Substituting the above values into Eq.~(\ref{eq:ene_reduce}), we find
\beq
\frac{N gE}{2\pi \mu} \!\!&=&\!\! \int^L_0\!\! dr\ r\bigg[
\frac{\hbar^2}{2m}\frac{r^2 + \lambda^2e^{-a r^2} + \left(a r^2+1\right)^2\lambda^2e^{-a r^2} }{ \left(r^2 +\lambda ^2e^{-a r^2}\right)^2}\non
&&+\ \omega  \bigg(\frac{N \lambda ^2e^{-a r^2}}{r^2+ \lambda ^2e^{-a r^2}}-(N-1)\bigg)\bigg],
\eeq
where $L$ denotes the system size.
We next find the values of $\lambda$ and $a$ 
for minimizing the energy and read the size
from
$r = \lambda e^{-\frac{a}{2}r^2}$.
In our study, we observed complete agreement between 
the radii obtained by the numerical solutions 
and the variational method,
as shown in Fig.~\ref{fig:sizes}.

{\it Vortex molecules} ---
Next, let us consider the miscible case $g > \tilde g$ to observe non-axisymmetric
configurations, namely, the vortex molecules.

{\bf Dimers}: 
Vortex dimers are molecules of two vortices appearing in
two-component BECs \cite{Kasamatsu:2004}. 
The dimer is the simplest molecule that does not
have any degeneracies or substructures.

\begin{figure}[ht]
\begin{center}
\includegraphics[width=0.7\hsize]{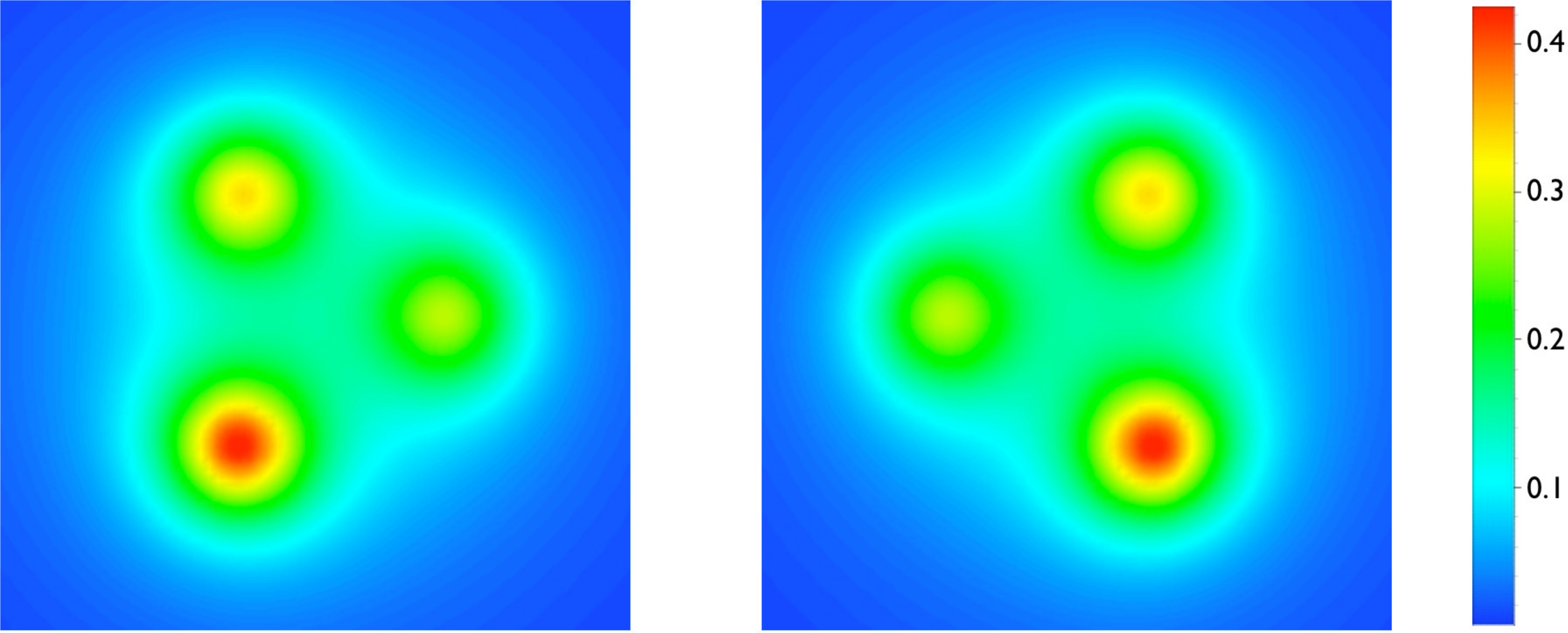}
\caption{Chirality: two trimers in a three-component BEC ($g_{11} \neq g_{22} \neq g_{33} \neq g_{11}$).
The plotted energy densities are shown in the figures. 
In each figure, 
there are three different vortices 
forming a molecule.
The left panel is a mirror image of the right panel.
The numerical range for each of the figures 
is set as $x \in [-1.5,1.5]$ and $y \in [-1.5,1.5]$.}
\label{fig:chiral}
\end{center}
\end{figure}

{\bf Trimers}:
Vortex trimers are molecules of three vortices appearing 
in three-component BECs. 
These exhibit two new properties that the dimers do not have.
The first reported in Ref.~\cite{Eto:2012rc}  
is that the shape of the triangle can be changed
by tuning the parameters in Eq.~(\ref{eq:gp}).
The second is the chirality of the triangle. 
This can be easily seen 
when the coupling constants are generic, 
so that the vortices have different sizes,  
as seen in Fig.~\ref{fig:chiral}.
In two-component BECs, chirality does not make sense because 
left and right can be exchanged by $\pi$ rotation.
On the other hand, as shown in Fig.~\ref{fig:chiral}, 
the left 
and right configurations 
cannot be transformed to each other by 
a rotation 
but can be transformed by a mirror.
The trimers are energetically completely degenerate.
\begin{figure*}
\begin{center}
\includegraphics[width=\hsize]{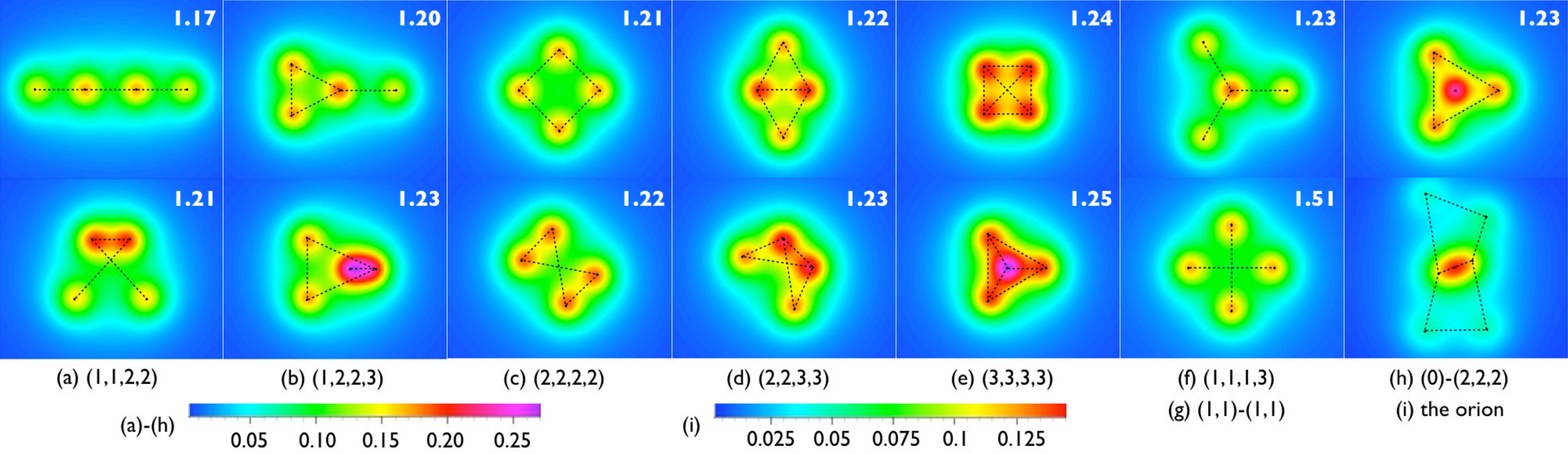}
\caption{(a)--(h) All possible tetramers in a four-component BEC 
and (i) the orion in a seven-component BEC. 
We set $\tilde g=900$ while 
the other parameters are the same as those for Fig.~\ref{fig:axisymmetric}.
For the tetramers,  we set
$\omega_{ij}=0.05$ and $\omega_{ij} = 0$ for
the connected and disconnected pairs, respectively.
For the orion, we set $(\omega_{12},\omega_{23},\omega_{34},\omega_{45},\omega_{16},\omega_{56},\omega_{37},\omega_{67})=
(0.02,0.001,0.02,0.02,0.05,0.02,0.2,0.2)$ and the rest to be zero.
The dots and dotted lines indicate the vortex center of each component and 
the Rabi couplings between components, respectively. 
The color maps represent the energy density 
(the left for tetramers and the right for the orion). 
The number at the upper-right corner in each panel 
represents the total energy. 
The size of the simulation box is $(2L)^2 = 30^2$ with the lattice space set to be less than 0.2.
The numerical range for each of the figures (a)--(h) 
is set as $x \in [-2.5,2.5]$ and $y \in [-2,2]$ and the one for (i) as $x \in [-3.125,3.125]$ and $y \in [-2.5,2.5]$.
}
\label{fig:4comp}
\end{center}
\end{figure*}

{\bf Tetramers}:
In multi-component BECs with more than four condensate wave functions, 
the number of possible molecules increases exponentially. 
Therefore, 
in order to classify such vortex molecules, 
we make use of the mathematical graph theory, 
in which vortices are expressed by vertices and 
the Rabi couplings are expressed by edges. 
Graphs isomorphic to each other are not distinguished 
in the graph theory. 
The number of independent connected graphs 
relevant for four-component BECs
is six. 
In the graph theory, a graph is characterized by the sequence of 
the number of edges connected to each vertex. 
For instance, $(1,1,2,2)$ implies that two vertices are connected by two edges, and other two vertices are connected by one edge.
We obtained numerical solutions 
for a four-component BEC with the same values of Rabi couplings 
(Fig.~\ref{fig:4comp}),
exhausting all possible graphs 
with four vortices as vertices 
including six connected graphs seen in Fig.~\ref{fig:4comp} (a)--(f) 
and two disconnected graphs in Fig.~\ref{fig:4comp} (g) and (h).
We solved the GP equation by the imaginary time propagation
with various initial conditions in which fractional vortex in each component 
is placed at various points, namely $\psi_i(t=0) \sim z - z_i$.
We always reached 
the same configurations (up to rotation), 
and therefore, we conclude that the vortex molecules are (meta)stable.
In order to clarify which configuration is stable at 
the global minimum or metastable at local minima,
we calculated the GP energy (\ref{eq:gp})
subtracted by the ground state energy for each configuration, 
as shown at the top-right of 
each panel in Fig.~\ref{fig:4comp}, where we chose 
$r=6$ for the spatial integration in Eq.~(\ref{eq:gp}). 

Our simulation indicate the occurrence of several new phenomena 
that do not exist in dimers or trimers:
1) twist, 2) holding, and 3) capture.
We have found that most molecules are accompanied by 
``twisted'' molecules, as in Fig.~\ref{fig:4comp} (a)--(e). 
A pair of untwisted and twisted molecules corresponds 
to identical graphs isomorphic to each other, 
while 
both pairs are energetically stable, corresponding to
absolute and local minima; 
the upper configuration has lesser energy than 
the lower configuration 
in each figure in Fig.~\ref{fig:4comp} (a)--(e),
and corresponds to the absolute minimum.
There exists an energy barrier between twisted and untwisted molecules, 
and therefore,  
once a twisted vortex molecule is formed, 
a certain amount of finite energy is required to ``untwist" it. 
As an example, a rod molecule denoted by 
the graph $(1,1,2,2)$ in 
(a) 
has an energy value of 
$1.17$ 
when it is straight, while its energy value is 
$1.21$ when it is twisted.

The second and third new phenomena can be observed 
in the disconnected graphs (g) and (h). 
The holding phenomenon can be observed in Fig.~\ref{fig:4comp} (g).
This is seen as the union of the two graphs with $(1,1)$.
If the two domain walls connecting vortices are not crossed, 
this molecule breaks up into two dimers 
because of repulsion between the vortices. 
The third phenomenon observed is the absorption 
of a molecule inside 
a bigger molecule, as in Fig.~\ref{fig:4comp} (h).

In the mathematical graph theory, 
an energy component called 
the graph energy is defined motivated by the 
$\pi$-electron energy of molecules in chemistry
\cite{graph-energy}.  
In such a case, for a given graph, 
configurations with the minimum graph energy 
are unique.
Our GP energy (\ref{eq:gp}) provides 
a new definition of graph energy in 
the graph theory, 
which makes vortex molecules 
more interesting even mathematically.

{\bf $N$-omers}:
We can engineer as many multiple vortex molecules as we want with the constituent vortices. 

Thus far, we have concentrated on the case when 
$\omega_{ij}=\omega \neq 0$ and $\omega_{ij} = 0$  
for the connected and disconnected pair of the vortices,
respectively.
We can control the positions and shape of the molecule 
by varying $\omega_{ij}$ inhomogeneously. 
As an example, we present a vortex heptamer (seven vortices), 
designed as the orion in Fig.~\ref{fig:4comp} (i).

\begin{figure}[ht]
\begin{center}
\includegraphics[width=8cm]{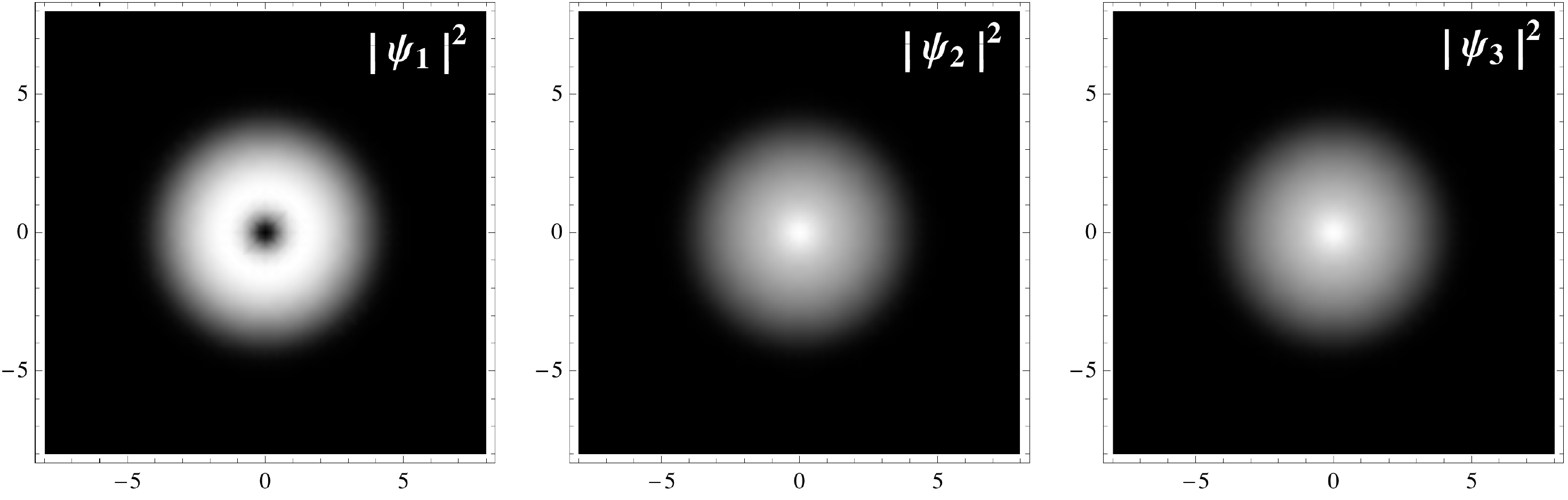}
\caption{
A minimal fractional vortex which has a non-trivial winding for the first condensate $\Psi_1$. 
The densities are $|\Psi_1|^2$, $|\Psi_2|^2$ and $|\Psi_3|^2$, 
where a black dot denotes a vortex. 
The parameters are $g_{ii}=1$, $g_{ij} =1/2$ for $i \neq j$, $\mu_i=10$, $\omega_i=1$ and $\Omega=0.1$.
}
\label{fig:fractional_vortex}
\end{center}
\end{figure}

\begin{figure}[ht]
\begin{center}
\includegraphics[width=8cm]{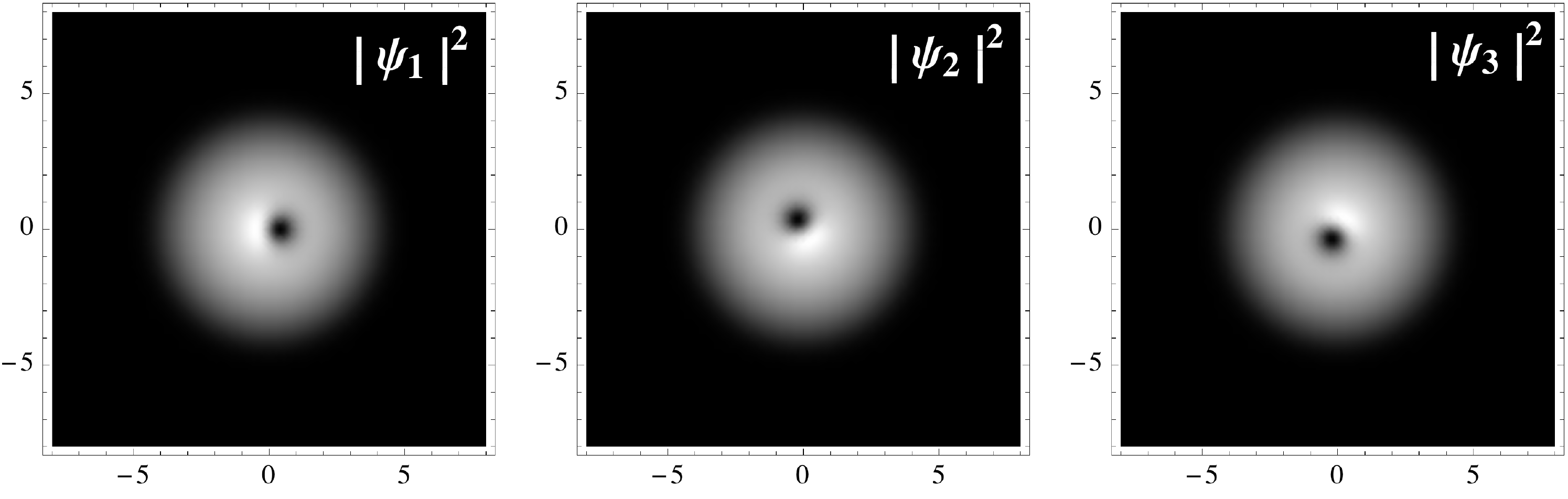}\\
\includegraphics[width=8cm]{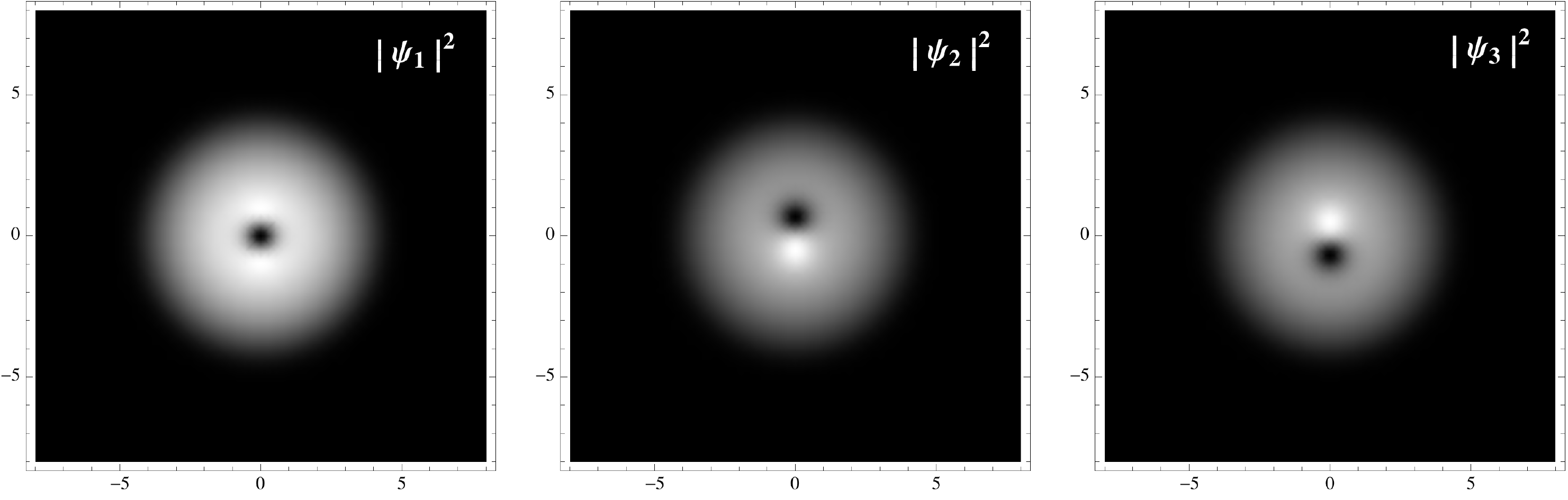}
\caption{
The vortex trimers in the $N=3$ case. The triangle trimer is shown on the first line with
$\omega_{12} = \omega_{23} = \omega_{31} = 0.1$ and the rod trimer is on the second line
with $\omega_{12} = \omega_{31} = 0.1$.
The convention of the tone is the same as that of Fig.~\ref{fig:fractional_vortex}. 
}
\label{fig:trime_rotate}
\end{center}
\end{figure}

\begin{figure}[ht]
\begin{center}
\includegraphics[width=8cm]{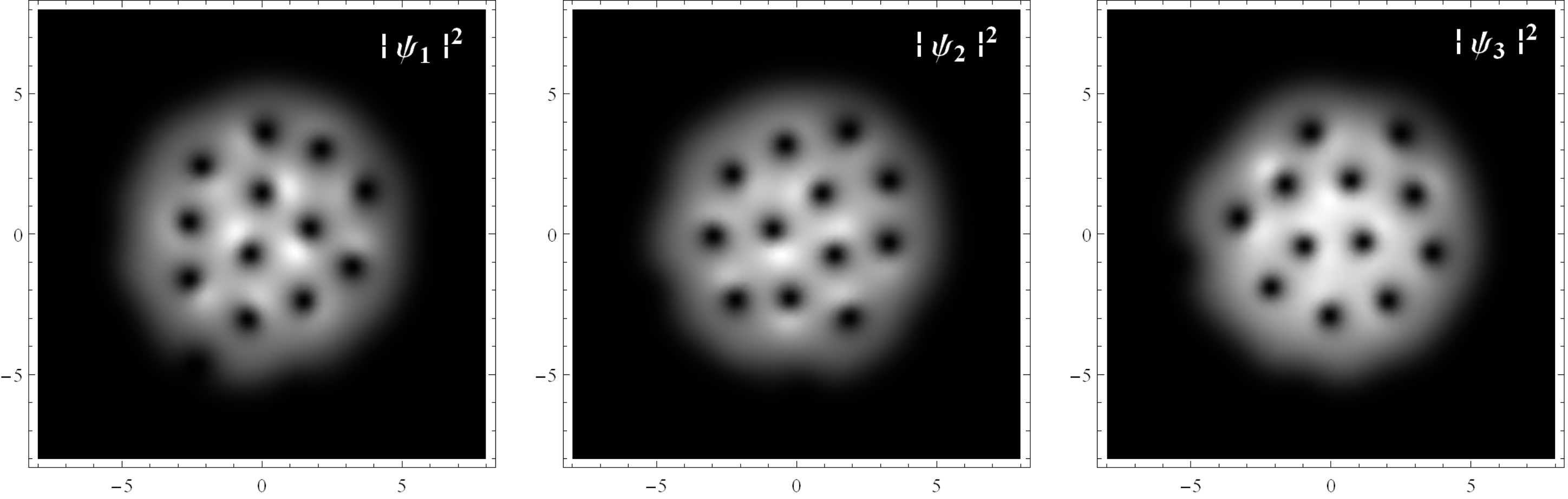}\\
\includegraphics[width=4cm]{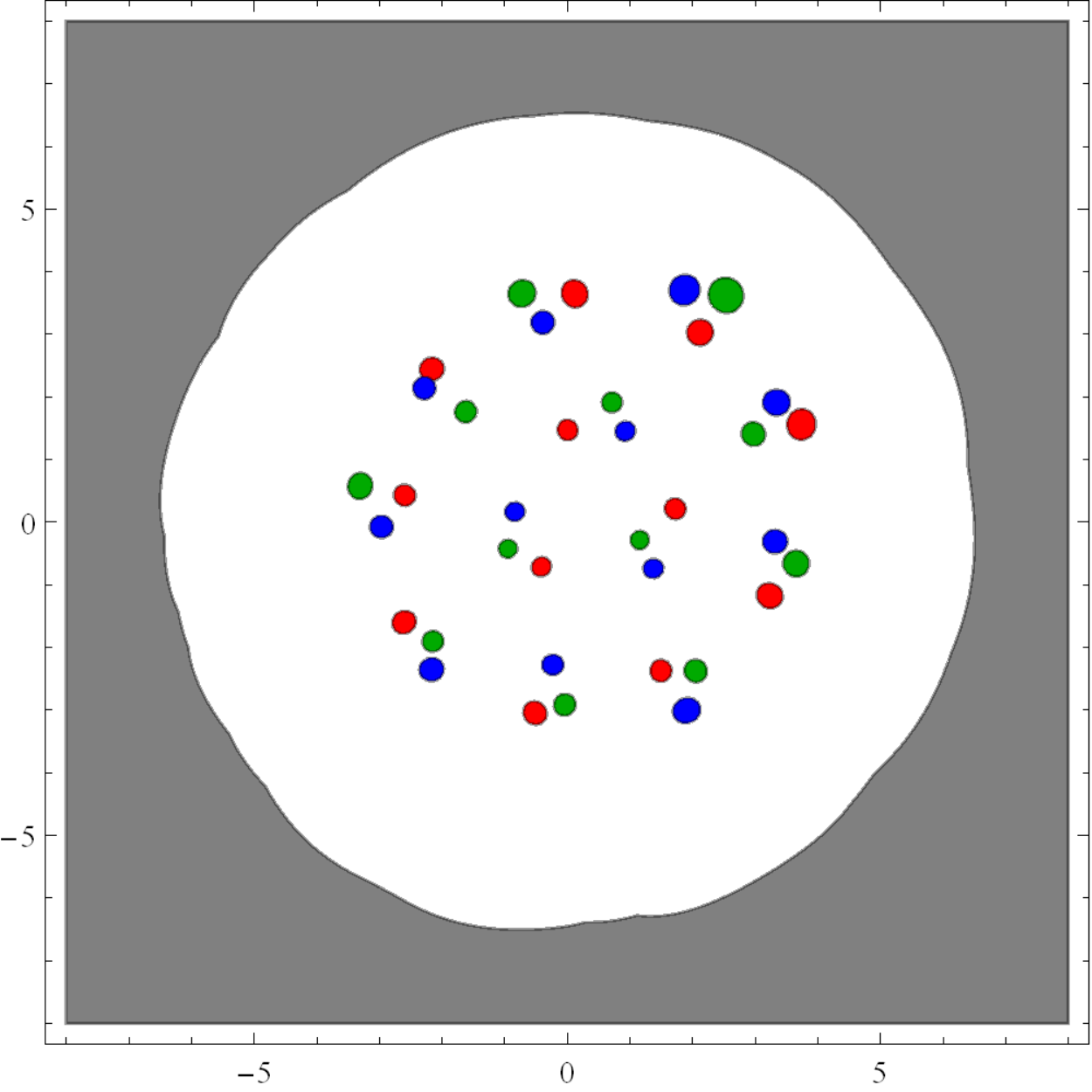}
\caption{
A lattice of vortex trimers. There exist two type of trimers with different chirality.
The parameters are the same as those in Fig.~\ref{fig:fractional_vortex} except for $\Omega=0.7$
and $\omega_{ij}=0.1$.
The convention of the tone is the same as that of Fig.~\ref{fig:fractional_vortex}. 
On the second line, we plot the superimposition of 
the three components where colors 
represent vortices in each component. 
}
\label{fig:lattice}
\end{center}
\end{figure}

{\it Vortex molecules in a rotating system} ---
The above solutions in non-rotating BECs with the Dirichlet boundary conditions imply the existence of 
the vortex $N$-omers in rotating BECs with 
a trapping potential. 
In order to see this, we further study formation of the molecules in a rotating system with a harmonic trapping potential, described by 
the GP equations 
\beq
\left(-\frac{\hbar^2}{2m_i}\nabla^2 + V_i-\mu_i +  \sum_{j=1}^N g_{ij}|\Psi_j|^2
-\Omega L_z \right)\Psi_i = 0,
\label{eq:GP_rotate}
\eeq
with a harmonic potential $V_i = \omega_i^2 r^2/2$ and a rotating angular velocity $\Omega$.
For simplicity, we solve the GP equation (\ref{eq:GP_rotate}) by imaginary time propagation for 
the $N=3$ case in what follows. 
First, we confirm the stability of 
a minimally winding fractional
vortex without the Rabi interactions,  as shown in Fig.~\ref{fig:fractional_vortex}.
We also have found a stable vortex trimer  
with the Rabi interactions   
and a rotation speed for a unit circulation,   
as shown Fig.~\ref{fig:trime_rotate}.
To obtain this configuration, 
there are two different ways: 
the Rabi coupling can be turned on either  
from the beginning or after three fractional vortices 
are formed. 
We did not find metastable states 
other than a single vortex trimer 
in the case of the rotation speed for 
the unit circulation, 
unlike multiple vortices in 
a two-component BEC \cite{Ji:2008}. 
If we further increase the rotation speed, 
a lattice of vortex trimers start to form 
as in Fig.~\ref{fig:lattice},  
where we found trimers with different chiralities in Fig.~\ref{fig:chiral}.
(For a vortex lattice of a three-component BEC
without the Rabi interactions, see Ref.~\cite{Cipriani:2013wia}.)

In conclusion,
we have no qualitatively different configurations
between static molecule solutions in the ideal situation of an infinite plane system and
a more realistic situation with rotation and a trapping potential. Although there are some
quantitative difference, 
the topological structures are the same.

{\it Discussion} ---
In conclusion,
we have demonstrated the construction of stable-vortex $N$-omers 
in coherently coupled $N$-component BECs, and 
we have classified all possible $N$-omers in terms of graph theory. 
One graph may correspond to two stable configurations 
which are the local and global minima.
We observed several new phenomena including chirality, twist, holding, and capture. 
We also found that $N$-omers are well described by 
${\mathbb C}P^{N-1}$ skyrmions for the $U(N)$ symmetric couplings. 
We used various 
initial conditions and always reached 
the same configurations (up to rotation), 
and therefore, we conclude that our results are consistent and verifiable.
The GP energy provides a new definition of graph energy 
in the graph theory. 
When the values of $\omega_{ij}$'s are not equal, 
the domain walls have tensions different from each other 
and the corresponding edges 
have weights in the graph theory.
When $v_i$'s are not equal, the vortices have different 
masses and the corresponding vertices have weights. 

In experiments, an integer vortex is created as 
usual by gradually increasing 
the rotation, and subsequently, it is split into a vortex $N$-omer. 
Two-component BECs of different hyperfine states of 
the same atom have been already realized using
the $|1,-1\big>$ and $|2,1\big>$ states \cite{Matthews} 
and the $|2,1\big>$ and  $|2,2\big>$ states \cite{Maddaloni:2000} 
of $^{87}$Rb, respectively. 
We believe that systems with three or more components can be realized by using a mixture of the above mentioned states of $^{87}$Rb 
via an optical trap \cite{Hamner:2011}, 
and our prediction is testable in laboratory experiments.

\ 

This work is supported in part by 
KAKENHI (No. 23740226 (ME),
and 
No. 23740198, 
No. 25400268,
No. 23103515, 
and 
No. 25103720 (MN))


\end{document}